\documentclass[%
reprint,aps,amsmath,amssymb
]{revtex4-2}

\usepackage{color}
\usepackage{graphicx}% Include figure files
\usepackage{dcolumn}% Align table columns on decimal point
\usepackage{bm}% bold math
\usepackage[colorlinks=true,citecolor=blue,urlcolor=blue,linkcolor=blue,hyperindex]{hyperref}
\usepackage{mathtools}
\usepackage{float}
\usepackage{booktabs}
\usepackage{subfig}
\makeatother
\begin{document}

\preprint{APS/123-QED}

\title{The possible coexistence of superconductivity and topological electronic states in 1T-RhSeTe}

%\thanks{Thankes}%

\author{Tengdong Zhang$^{1}$}
\author{Rui Fan$^{3}$}%
\author{Yan Gao$^{1}$}
\author{Yanling Wu$^{1}$}
\author{Xiaodan Xu$^{1}$}
\author{Dao-Xin Yao$^{2}$}
\email{yaodaox@mail.sysu.edu.cn}
\author{Jun Li$^{1}$}
\email{ljcj007@ysu.edu.cn}
\affiliation{
$^1$Key Laboratory for Microstructural Material Physics of Hebei Province, School of Science, Yanshan University, Qinhuangdao 066004, China.\\
$^2$State Key Laboratory of Optoelectronic Materials and Technologies, Guangdong Provincial Key Laboratory of Magnetoelectric Physics and Devices, School of Physics, Sun Yat-Sen University, Guangzhou 510275, Peoples Republic of China.\\
$^3$School of Physics, Beijing Institute of Technology, Beijing 100081, China}

\begin{abstract}
Transition metal dichalcogenides (TMDs), exhibit a range of crystal structures and topological quantum states. The 1$T$ phase, in particular, shows promise for superconductivity driven by electron-phonon coupling, strain, pressure, and chemical doping. In this theoretical investigation, we explore 1$T$-RhSeTe as a novel type of TMD superconductor with topological electronic states. The optimal doping structure and atomic arrangement of 1$T$-RhSeTe are constructed. Phonon calculations validate the integrity of the constructed doping structure. The analysis of the electron-phonon coupling (EPC) using the Electron-phonon Wannier (EPW) method has confirmed the existence of a robust electron-phonon interaction in 1$T$-RhSeTe, resulting in total EPC constant $\lambda$ = 2.02, the logarithmic average frequency $\omega_{\text{log}}$ = 3.15 meV and $T_c$ = 4.61 K, consistent with experimental measurements and indicative of its classification as a BCS superconductor. The band structure analysis revealed the presence of Dirac-like band crossing points. The topological non-trivial electronic structures of the 1$T$-RhSeTe are confirmed via the evolution of Wannier charge centers (WCCs). Collectively, these distinctive properties underscore 1$T$-RhSeTe as a possible candidate for a topological superconductor, warranting further investigation into its potential implications and applications.

\end{abstract}
\maketitle

\section{\label{sec:intro}INTRODUCTION}

\begin{figure*}[t]
	\centering
	\includegraphics[width=0.95\textwidth]{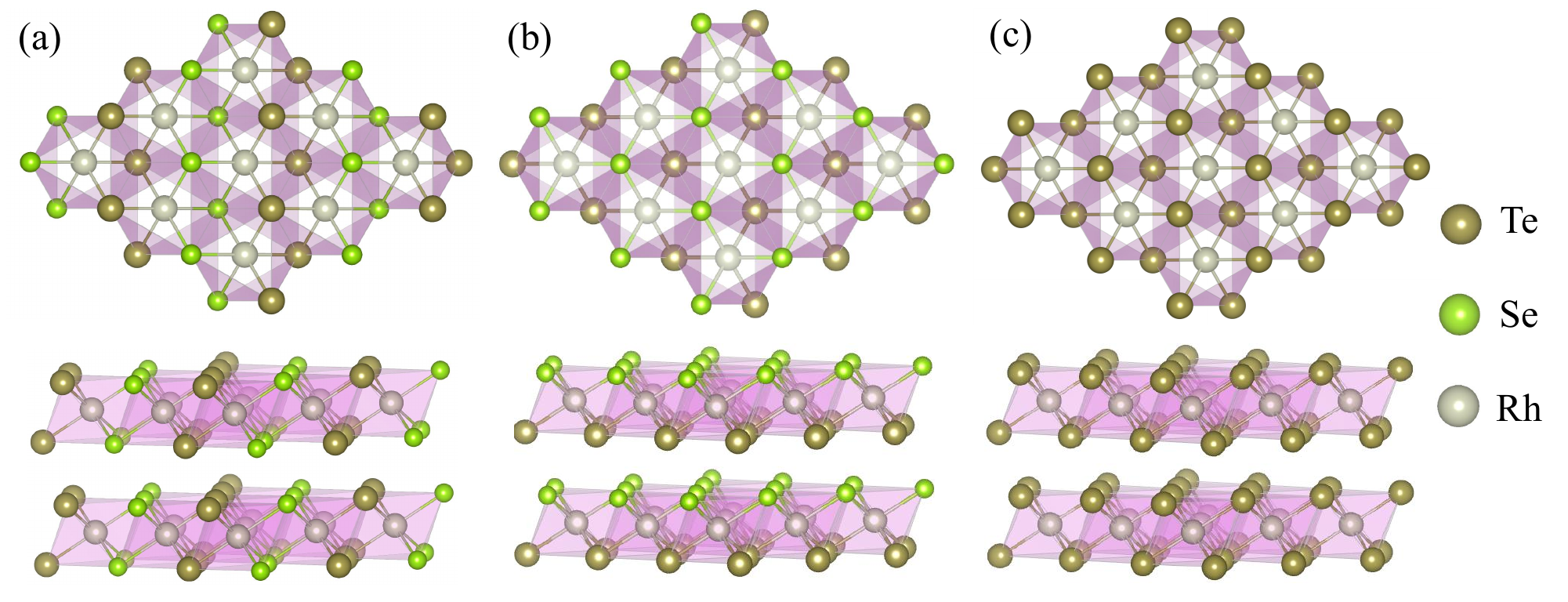}
	\caption{\label{cif}Top and side views of the crystal structure of 1$T$-RhSeTe and RhTe$_2$: (a) type-I doping 1$T$-RhSeTe, where Se replaces Te atoms along lines; (b) type-II doping 1$T$-RhSeTe, where Se replaces Te atoms within plane; (c) 1$T$-RhTe$_2$.}
\end{figure*}

Transition metal dichalcogenides (TMDs) are compounds represented by the chemical formula MX$_2$, where M denotes transition metals such as Mo, W, and Ni, and X represents chalcogen elements such as S, Se, and Te. Featuring a layered structure, TMDs display a remarkable degree of diversity and exhibit a variety of topological quantum states, including Weyl semimetallic, topological insulator, and quantum spin Hall effect \cite{di_sante_three-dimensional_2017,Wu2019,Wu2018,Costanzo2016}. TMDs can adopt various crystal structures, with the most prevalent being hexagonal (2-$H$), rhombohedral (3-$R$), and trigonal (1-$T$) phases. The 2-$H$ phase typically manifests as a semiconductor and is the most thermodynamically stable, while the 1-$T$ phase exhibits metallic behavior and possesses distinctive topological properties \cite{Fang2018,Bruyer2016,Strachan2021}. 

The superconducting behaviors of 1-$T$ phase TMDs are intrinsic or induced by chemical doping or applied strain \cite{Lian2021,He2018,Zhao2020}. For example, both 1$T$-TiSe$_2$ and 1$T$-PdTe$_2$ exhibit semimetallic characteristics, and the topological superconductivity can be driven by electron-phonon coupling. Experimental observations have revealed that 1$T$-TiSe$_2$ exhibits a critical temperature ($T_c$) of 2 K under a pressure of 3 GPa, while the $T_c$ of PdTe$_2$ is 1.7 K \cite{Kusmartseva2009,Xiao2017}. In the 1-$T$ phases of PtTe$_2$, PdSe$_2$, NiTe$_2$, and PdTe$_2$, the Dirac cone is tilted, and forms a type-II Dirac topological point \cite{He2014,Ermolaev2022,Yan2017,Tanisha2024}. Furthermore, the Dirac point in NiTe$_2$ is situated at the Fermi energy level of 0.80 meV, and  the position of the Fermi energy can be adjusted through strain, pressure, and chemical doping \cite{Zhao2020}. 1$T$-phase TMDs hold significant promise for inducing superconductivity with topological quantum states. 

In the broadest sense, a superconductor is regarded as topological if any of the topological numbers is nonzero \cite{Sato2017,Gu2020}. Topological superconductors have emerged as a fascinating area of research in quantum computing, due to their unique physical properties and potential applications \cite{Zhang2020,Kitaev2003, Lahtinen2017,Nayak2008,Sarma2015}. They offer an ideal platform for studying fundamental quantum physics problems, such as non-abelian anyons and quantum entanglement, and propose possible solutions for building future quantum computers that can achieve fault-tolerant quantum computation \cite{Nayak2008,Fidkowski2010,Sarma2015}. Topological superconductors are closely associated with Majorana fermions, which are exotic particles that are their own antiparticles \cite{Qi2011,Sarma2015}. 

A novel Dirac semimetal, 1$T$-RhSeTe, has been identified as a superconductor with $T_c$ = 4.72 K \cite{Patra2023}. In 1$T$-RhSeTe, selenium (Se) replaces tellurium (Te) atoms in the 1$T$ phase of RhTe$_2$, a TMD previously reported to lack superconductivity and topological electronic states \cite{Geller1955,Lurgo2022}. This substitution induces superconductivity and topological electronic states, which is a possible candidate for topological superconductors.

This study employs density-functional theory (DFT) within first-principles calculations to examine the electronic bands, phonon, superconductivity, and topological characteristics of 1$T$-RhSeTe. Analysis of the electron-phonon coupling and Eliashberg spectral function yields the electron-phonon coupling strength ($\lambda$ = 2.02 ) and $T_{c}$ = 4.61 K.

Furthermore, a Dirac cone crosses near the Fermi energy in the absence of SOC. However, the presence of the SOC effect result in the Dirac cone becoming gapped, thereby manifesting a Dirac-type surface state that is protected by time-reversal symmetry. The calculated $\mathbb{Z}_2$ topological index is $\mathbb{Z}_2$ = 1, indicating that 1$T$-RhSeTe is in a strongly topological electronic state.

\section{\label{sec:mod_and_approa}Computational Details}

\begin{table}
    \caption{\label{tab:my_label}Lattice constants for type-I, type-II 1$T$-RhSeTe, compare with experiment and  1$T$-RhTe$_2$}
    \begin{tabular}{l|lll}
    \hline \hline
      Structure & Lattices(\AA) & Energy & Ref.  \\ \hline
        Type-I  & a, b = 3.794, c = 5.363 &  -32.036 eV &calculate\\
        Type-II & a, b = 3.818, c = 5.291 &  -32.012 eV &calculate\\
        Exper.  & a, b = 3.798, c = 5.389 &   & Ref.\cite{Patra2023}\\
        RhTe$_2$& a, b = 3.938, c = 5.513 &   & 
        Ref.\cite{Geller1955}\\ \hline \hline
    \end{tabular}
\end{table}

We use the projector-augmented wave method implemented in the Vienna Ab initio Simulation Package (VASP) to perform density-functional theory (DFT) calculations for the structure optimization \cite{Kresse1993}. The generalized gradient approximation (GGA) and the Perdew Burke-Ernzerhof (PBE) function are used to treat the electron exchange-correlation potential\cite{Perdew1996,Kresse1999}. In the computation, the plane wave cut-off energy is set at 700 eV, a Monkhorst-Pack k lattice with a spacing of 2$\pi\times$ 0.03 \AA$^{-1}$ is employed, and the structure is optimized using optB88-vdW van der Waals (vdW) interaction corrections \cite{Klimes2009} until the force on each atom less than 0.015 eV$ \times$ \AA$^{-1}$.

Quantum Espresso (QE) is used for DFT calculations including self-consistent, band structure, and phonon calculations \cite{Giannozzi2017}. The PBE generalization of GGA is the basis for the exchange-correlation generalization. The calculations used kinetic energy cutoff values of 70 Ry for the wave function and 700 Ry for the potential function, a 6 $\times$ 6 $\times$ 6 k-grid, and a 6 $\times$ 6 $\times$ 6 q-point grid. The self-consistent convergence threshold is set to 1 $\times$ 10$^{-10}$ Ry. The d-orbitals of the Rh atoms, the p-orbitals of the Te atoms, and the p-orbitals of the Se atoms are selected as the projected orbits, and the 6 $\times$ 6 $\times$ 6 k-mesh grid is set up to construct the tight-binding Hamiltonian for the established structure by the software package wannier90 based on maximally localized wannier functions (MLWFs) \cite{Mostofi2014,Mostofi2008}.

The Electron-phonon Wannier (EPW) code is employed to obtain the electron-phonon coupling and BCS superconductivity. The k-mesh and q-mesh grids are uniformly increased to 20 $\times$ 20 $\times$ 20 for more accurate results. The EPW code solves the Eliashberg equation and related superconductivity equations to calculate electron-phonon coupling strength and $T_c$. The electron-phonon coupling constant $\lambda$, which can be calculated by  
\begin{equation}\label{eq-lambda}
    \lambda=2\int \frac{\alpha^{2}\left ( \omega  \right )  }{\omega } d\omega, 
\end{equation}
where $\alpha^2F(\Omega)$ is the Eliashberg spectral function, the phonon density of states weighted by an electron-phonon coupling matrix element at each k-point and frequency. The superconducting $T_c$ is determined with the McMillan equation modified by Allen-Dynes formula \cite{allen_neutron_1972,Allen1975} 
\begin{equation}\label{eq-tc}
    T_c=\frac{f_1f_2\omega_{log} }{1.2} \mathrm{exp}\left [ \frac{-1.04\left ( 1+\lambda  \right ) }{\lambda \left ( 1-0.62\mu ^{*}  \right )-\mu^{*}  }  \right ], 
\end{equation}
where $\mu^*$ is the effective screened Coulomb repulsion constant typically ranging from 0.1 to 0.16 eV, and the logarithmically averaged frequency $\omega_{log}$ can be defined as
\begin{equation}
    \omega_{log}=\mathrm{exp} \left [ \frac{2}{\lambda } \int \frac{d\omega }{\omega}\alpha ^{2}F(\omega)\ln{\omega} \right ].
\end{equation}

We use WannierTools\cite{Wu2018a} to get WCCs \cite{Yu2011} to reflect the topological properties of time-reversal invariant systems. $\mathbb{Z}_2$ topological invariant \cite{Kane2005} is given by the number of times $L$ any line %a/c($k_i$)
crosses a WCCs line when going from $k_i$ = 0 to $k_i$ = $\pi$. 
\begin{equation}
\mathbb{Z}_2 = L\mod 2
\end{equation}
The topological properties of three-dimensional materials can be classified using the $\mathbb{Z}_2$ topological index, classified by a set of indices $(\nu;\, \nu_x\, \nu_y\,\nu_z)$ defined through the 2D invariants on the time-reversal invariant planes in the Brillouin zone
\begin{equation}\label{eq-nu}
    \nu=\mathbb{Z}_2(k_i=0)+\mathbb{Z}_2(k_i=0.5)\mod 2 
\end{equation}
\begin{equation}
    \nu_i=\mathbb{Z}_2(k_i=0.5)
\end{equation}
where $k_i$ is in reduced coordinates. A system is referred to as a strong topological electronic state if $\nu$ = 1. 

\section{\label{sec:sc}results and discussion}

\begin{figure}[]
	\centering
	\includegraphics[width=0.5\textwidth]{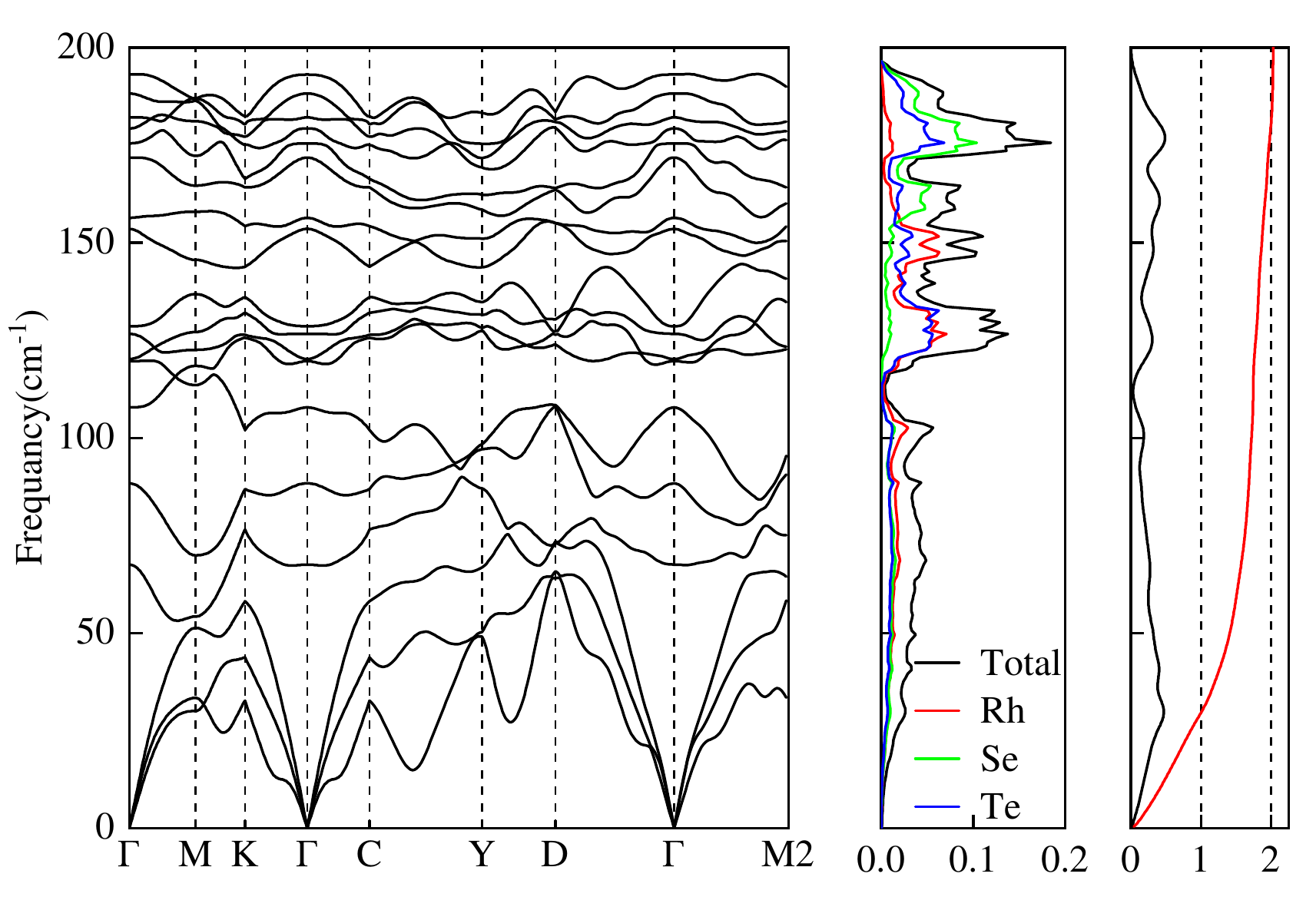}
	\caption{\label{epc}The phonon spectrum of the 1$T$-RhSeTe, phonon density of states (PHDOS), and Eliashberg spectral function $\alpha^2F(\omega)$.}
\end{figure}

\begin{figure}[]
\centering
{\includegraphics[width=1.0\linewidth]{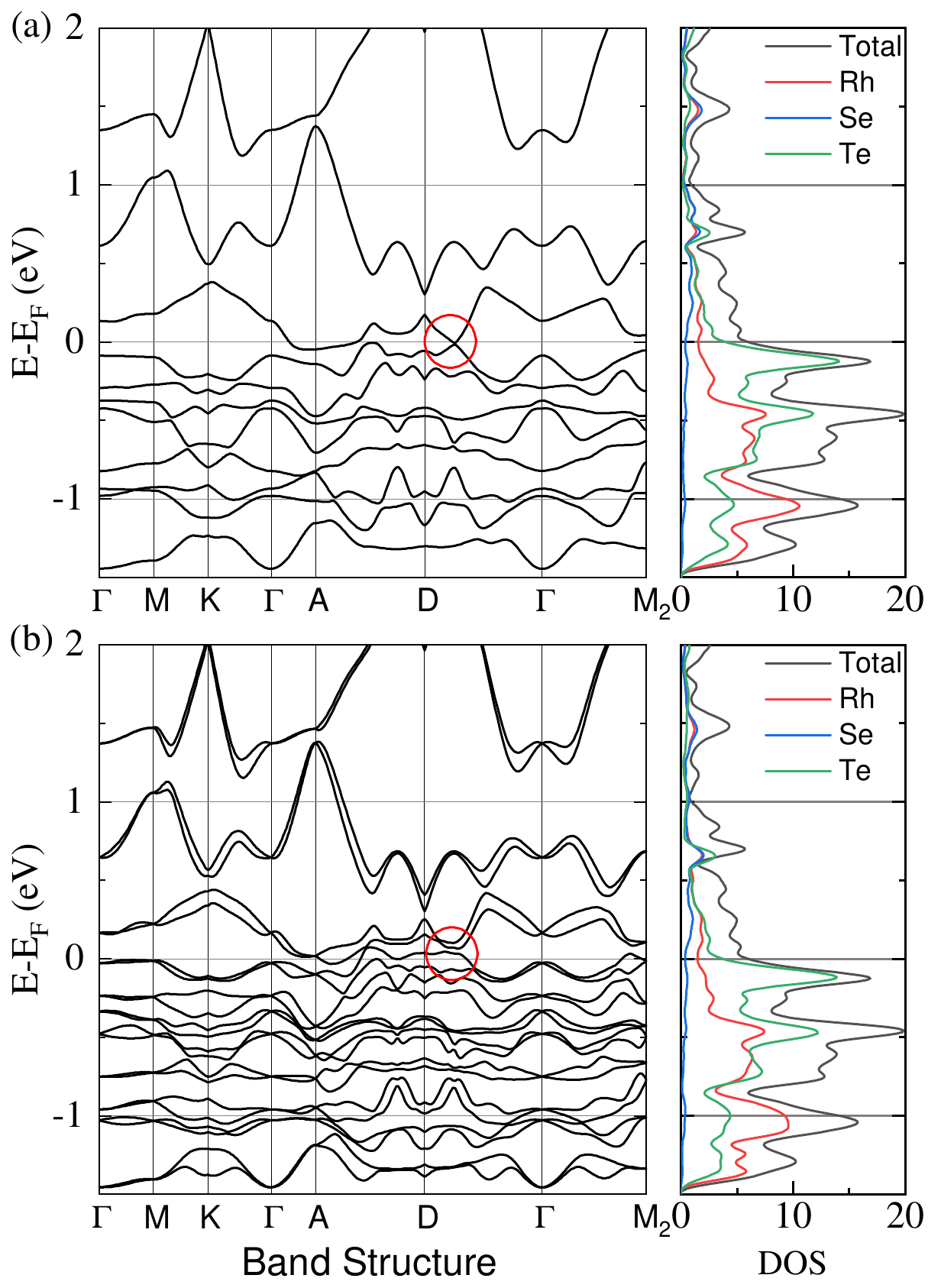}}
\caption{\label{fig1a} Band structure and density of states of the 1$T$-RhSeTe (a) without SOC and (b)including SOC obtained from DFT calculations. There is a Dirac point between $D$ and $\Gamma$. }
\end{figure}

The 1$T$-RhSeTe sample obtained in the experiment was prepared by doping Te with Se, with an exact stoichiometry of Rh$_{1.01}$Se$_{1.05}$Te$_{0.94}$. Two potential doping structures, designated as type-I and type-II, are illustrated in Fig. \ref{cif}(a) and (b) respectively. In the Type-I doping of 1$T$-RhSeTe, Te atoms are substituted by Se atoms in a linear configuration, resulting in a zigzag arrangement of Te and Se atoms. This arrangement disrupts the inherent triple rotational symmetry of 1$T$-RhTe$_2$ and leads to a slightly shorter Se-Rh bond compared to the Te-Rh bond, thereby reducing the symmetry to P2$_1$/m (No. 11). In contrast, Type-II doping 1$T$-RhSeTe involves the replacement of Te atoms by Se atoms within a plane, where Te and Se atoms are distributed in planes, disrupting only the pre-existing -3 symmetry and reducing the symmetry to P3m1 (No. 156). The symmetry of Type-I doped 1$T$-RhSeTe, which is P2$_1$/m (No. 11), is lower than that of Type-II doping 1$T$-RhSeTe [P3m1 (No. 156)]. Both of these symmetries are lower than the P-3m1 (No. 164) symmetry of 1$T$-RhTe$_2$. 

The lattice constants of the two structures following structural relaxation are presented in Tab \ref{tab:my_label}. Type-II doping of 1$T$-RhSeTe occupies a larger portion of the lattice within the plane, and the two different lengths of Se-Rh and Te-Rh bonds are forced to twist onto the same plane, resulting in a higher energy state compared to Type-I of doping 1$T$-RhSeTe. Significantly, the lattice constants of Type-I doping 1$T$-RhSeTe are more closely aligned with experimental measurements, suggesting that the experimental measured 1T-RhSeTe is more likely to conform to Type-I doping. Consequently, we only consider the Type-I doping structure of 1$T$-RhSeTe in the subsequent analysis.

Phonon dispersion represents a fundamental approach to assessing the the dynamic stability of a system. The calculated phonon dispersions of 1$T$-RhSeTe are presented in Fig. \ref{epc}. The absence of imaginary phonon modes in 1T-RhSeTe is indicative of its dynamic stability. The phonon dispersion can be classified into two distinct regions based on the vibration frequency. In the high-frequency region ($\omega$ $\geq$ 110 cm $^{-1}$), the majority of the phonons are composed of the vibration modes of Se and Te atoms. The Se atoms, which have the lowest mass, contribute the most to the higher-frequency phonons. The Te atoms engage in a seesaw-like resonance around the neighboring Rh atoms with Se atoms, thereby contributing to an additional portion of high-frequency phonons. In the low-frequency region, the phonons are formed by a combination of the vibrational modes of Rh and Te atoms, as these atoms are heavier and more localized in frequency space than Se atoms. 

The calculated Eliashberg spectral and integrated electron-phonon coupling strength are plotted in the right panel of Fig. \ref{epc}. The distribution of the Eliashberg spectral function,$\alpha^2F(\Omega)$, in the high-frequency region is almost the same as in the low-frequency region. Notably, the vibration frequencies of the heavier elements, including Rh, Se, and Te, predominantly fall within the low-frequency range of 0-200 cm$^{-1}$. This behavior mirrors findings from electron-phonon coupling calculations in heavier element superconductors, such as LaBi$_3$ and LaCuSb$_2$, where electron-phonon interactions are concentrated in the lower vibration frequency range\cite{Chen_2023,akiba2023phonon}.

The electron-phonon coupling in the low-frequency region peaks at around $\omega$ $\sim$ 25 cm $^{-1}$, with obvious in-plane vibration modes around the M-K line. As indicated by Eq. (1), these in-plane vibration modes can result in a more pronounced electron-phonon coupling effect, thereby enhancing superconductivity. Around $\omega$ $\sim$ 180 cm $^{-1}$ in the high-frequency region, there is another electron-phonon coupling peak, which is mainly dominated by the seesaw-like vibration modes of Se and Te atoms. 

Utilizing MLWFs, we derive the isotropic Eliashberg spectral function $\alpha^2F(\omega)$ for 1$T$-RhSeTe. The integration of $\alpha^2F(\omega)$ yields the total EPC constant $\lambda$ = 2.02 and the logarithmic average frequency $\omega_{\text{log}}$ = 3.15 meV, indicating a strong electron-phonon coupling in the material. $T_c$ is estimated using the modified McMillan equation by Allen and Dynes. We select $\mu^*$ = 0.16 eV considering the heavy elements Rh, Se, and Te. Consequently, we obtain $T_c$ = 4.61 K, which is remarkably close to the experimental value of 4.72 K \cite{Patra2023}. This calculation lends credence to the system's classification as a BCS superconductor, bolstering the validity of our methods.

The energy band structure of 1$T$-RhSeTe is determined using QE, with and without SOC, as illustrated in Fig. \ref{fig1a}. In the absence of SOC, the four energy bands closest to the Fermi level are distinctly separated from the remaining bands, exhibiting clear differentiation. It is noteworthy that the two energy bands in closest proximity to the Fermi level exhibit a band crossing along the D-$\Gamma$ direction. This crossing point is located very close to the Fermi level (-0.01 eV) and exhibits a linear dispersion, resulting in the formation of Dirac fermions that are analogous to the Dirac points observed in graphene. The Fermi velocity at this point is $v_f$ = 1.71 $\times$ 10$^{5}$ ms$^{-1}$, which is much lower than that of graphene ($\sim$ $\times$ 10$^{6}$ ms$^{-1}$). This reduction in Fermi velocity can be attributed to the fact that these two bands are primarily derived from the p orbitals of Te. Given that the atomic mass of Te is significantly greater than that of C, the effective Fermi velocity is consequently much smaller than that observed in graphene.

\begin{figure}[]
	\centering
	\includegraphics[width=0.49\textwidth]{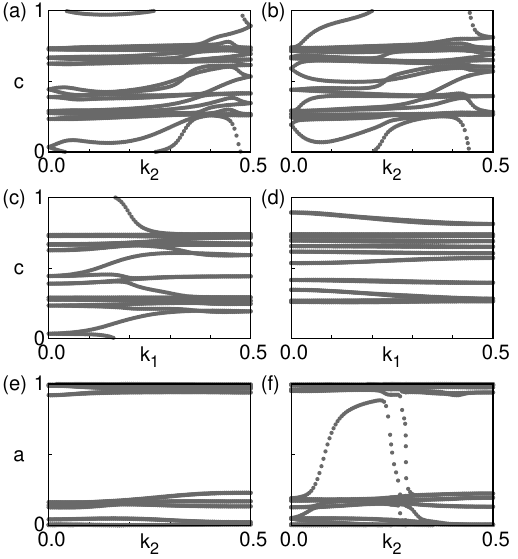}
	\caption{\label{z2}The evolution of WCCs in the six planes of in Brillouin Zone. The $\mathbb{Z}_2$
invariants are (a) $k_1$ = 0.0, $k_2-k_3$ plane: $\mathbb{Z}_2$ = 1 (b) $k_1$ = 0.5, $k_2-k_3$ plane: $\mathbb{Z}_2$ = 0 (c) $k_2$ = 0.0, $k_1-k_3$
plane: $\mathbb{Z}_2$ = 1 (d) $k_2$ = 0.5, $k_1-k_3$ plane: $\mathbb{Z}_2$ = 0 (e) $k_3$ = 0.0, $k_1-k_2$ plane: $\mathbb{Z}_2$ = 0 (f) $k_3$ = 0.5, $k_1-k_2$
plane: $\mathbb{Z}_2$ = 1.}
\end{figure}

A peak in the density of states (DOS) is observed approximately 0.1 eV below the Fermi level, predominantly contributed by the Te atoms. This van Hove singularity in the electron band, associated with the Te atoms, is beneficial for the formation of superconductivity. Additionally, it is noteworthy that this van Hove singularity originates from a flat band along the $\Gamma$-M direction. In conjunction with the aforementioned analysis of the low-frequency phonons contributed by Te atoms, it is proposed that this in-plane $\Gamma$-M flat band is pivotal for the emergence of superconductivity. Building upon the findings of previous on cuprate superconductors, we propose that electron doping or applying pressure could significantly enhance the superconductivity of 1$T$-RhSeTe, potentially approaching the McMillan limit.

In consideration of the SOC effect, the calculated band structure and DOS for 1$T$-RhSeTe are presented in Fig. \ref{fig1a}(b). In comparison to the band structures without the SOC effect, the band structures undergo significant changes due to the pronounced SOC resulting from the large atomic mass. Each energy band splits into two bands when including SOC, demonstrating that the 1$T$-RhSeTe system adheres to time-reversal symmetry. The energy and momentum positions of the Dirac cone in the $\Gamma$-D line are roughly 151 meV (another suspected Dirac cone at 115 MeV) above the Fermi energy level. The results of the calculation indicated the presence of topological non-trivial states in 1$T$-RhSeTe. 

FIG.\ref{z2} illustrates the evolution of WCCs in the six planes of the Brillouin Zone. Three planes exhibit a topologically nontrivial state, leading to the determination of the topological $\mathbb{Z}_2$ index for 1$T$-RhSeTe represented as (1 ; 0 0 1). The first $\mathbb{Z}_2$ index is the strong index, while the last three $\mathbb{Z}_2$ numbers are the weak indices. This topological $\mathbb{Z}_2$ index indicates that 1$T$-RhSeTe has a strongly topological electronic state.

\section{\label{sec:con}Conclusions}

In conclusion, theoretical calculations have been conducted on 1$T$-RhSeTe, which has been identified as a type of superconductor exhibiting topological electronic states. A doping structure and atomic arrangement for 1$T$-RhSeTe are constructed that optimizes the material's properties. The integrity of the constructed doping structure is validated through phonon calculations. The analysis of the electron-phonon coupling using the EPW method has confirmed the existence of a robust electron-phonon interaction in 1$T$-RhSeTe. This has led to the determination of total EPC constant $\lambda$ = 2.02, the logarithmic average frequency $\omega_{\text{log}}$ = 3.15 meV and $T_c$ = 4.61 K, consistent with experimental measurements and indicative of its classification as a BCS superconductor. The band structure analysis revealed the presence of Dirac-like band crossing points, van Hove singularities, and a flat band. The confirmation of the topological non-trivial electronic structures of 1$T$-RhSeTe is achieved through the evolution of WCCs. The coexistence of superconductivity and topological electronic states in 1$T$-RhSeTe indicates that it may be a promising candidate for topological superconductivity.

%There is potential to enhance both the topological and superconducting properties of 1$T$-RhSeTe. With regard to superconductivity, it can be posited that electron doping or the application of pressure could significantly enhance the superconducting properties of 1$T$-RhSeTe, given that the van Hove singularity and flat band in this material are in close proximity to the Fermi level. With regard to topological properties, this study considers 1$T$-RhSeTe as a three-dimensional material, identifying out-of-plane topological surface states. However, due to its inherent two-dimensional characteristics, it is also promising to explore the emergence of two-dimensional edge topological bands. 

\section{ACKNOWLEDGMENTS}
This work is supported by the National Natural Science Foundation of China (Grant No. 12204400), Science Research Project of Hebei Education Department (Grant No. QN2022169), Natural Science Foundation of Hebei Province (Grant No. A2022203010), Innovation Capability Improvement Project of Hebei province (Grant No. 22567605H).
T. Z. and R. F. contributed equally to this work. 
\nocite{*}
\bibliography{cite}% Produces the bibliography via BibTeX.

\providecommand{\noopsort}[1]{}\providecommand{\singleletter}[1]{#1}%
\begin{thebibliography}{44}%
\makeatletter
\providecommand \@ifxundefined [1]{%
 \@ifx{#1\undefined}
}%
\providecommand \@ifnum [1]{%
 \ifnum #1\expandafter \@firstoftwo
 \else \expandafter \@secondoftwo
 \fi
}%
\providecommand \@ifx [1]{%
 \ifx #1\expandafter \@firstoftwo
 \else \expandafter \@secondoftwo
 \fi
}%
\providecommand \natexlab [1]{#1}%
\providecommand \enquote  [1]{``#1''}%
\providecommand \bibnamefont  [1]{#1}%
\providecommand \bibfnamefont [1]{#1}%
\providecommand \citenamefont [1]{#1}%
\providecommand \href@noop [0]{\@secondoftwo}%
\providecommand \href [0]{\begingroup \@sanitize@url \@href}%
\providecommand \@href[1]{\@@startlink{#1}\@@href}%
\providecommand \@@href[1]{\endgroup#1\@@endlink}%
\providecommand \@sanitize@url [0]{\catcode `\\12\catcode `\$12\catcode
  `\&12\catcode `\#12\catcode `\^12\catcode `\_12\catcode `\%12\relax}%
\providecommand \@@startlink[1]{}%
\providecommand \@@endlink[0]{}%
\providecommand \url  [0]{\begingroup\@sanitize@url \@url }%
\providecommand \@url [1]{\endgroup\@href {#1}{\urlprefix }}%
\providecommand \urlprefix  [0]{URL }%
\providecommand \Eprint [0]{\href }%
\providecommand \doibase [0]{https://doi.org/}%
\providecommand \selectlanguage [0]{\@gobble}%
\providecommand \bibinfo  [0]{\@secondoftwo}%
\providecommand \bibfield  [0]{\@secondoftwo}%
\providecommand \translation [1]{[#1]}%
\providecommand \BibitemOpen [0]{}%
\providecommand \bibitemStop [0]{}%
\providecommand \bibitemNoStop [0]{.\EOS\space}%
\providecommand \EOS [0]{\spacefactor3000\relax}%
\providecommand \BibitemShut  [1]{\csname bibitem#1\endcsname}%
\let\auto@bib@innerbib\@empty
%</preamble>
\bibitem [{\citenamefont {Di~Sante}\ \emph {et~al.}(2017)\citenamefont
  {Di~Sante}, \citenamefont {Das}, \citenamefont {Bigi}, \citenamefont
  {Ergönenc}, \citenamefont {Gürtler}, \citenamefont {Krieger}, \citenamefont
  {Schmitt}, \citenamefont {Ali}, \citenamefont {Rossi}, \citenamefont
  {Thomale}, \citenamefont {Franchini}, \citenamefont {Picozzi}, \citenamefont
  {Fujii}, \citenamefont {Strocov}, \citenamefont {Sangiovanni}, \citenamefont
  {Vobornik}, \citenamefont {Cava},\ and\ \citenamefont
  {Panaccione}}]{di_sante_three-dimensional_2017}%
  \BibitemOpen
  \bibfield  {author} {\bibinfo {author} {\bibfnamefont {D.}~\bibnamefont
  {Di~Sante}}, \bibinfo {author} {\bibfnamefont {P.~K.}\ \bibnamefont {Das}},
  \bibinfo {author} {\bibfnamefont {C.}~\bibnamefont {Bigi}}, \bibinfo {author}
  {\bibfnamefont {Z.}~\bibnamefont {Ergönenc}}, \bibinfo {author}
  {\bibfnamefont {N.}~\bibnamefont {Gürtler}}, \bibinfo {author}
  {\bibfnamefont {J.~A.}\ \bibnamefont {Krieger}}, \bibinfo {author}
  {\bibfnamefont {T.}~\bibnamefont {Schmitt}}, \bibinfo {author} {\bibfnamefont
  {M.~N.}\ \bibnamefont {Ali}}, \bibinfo {author} {\bibfnamefont
  {G.}~\bibnamefont {Rossi}}, \bibinfo {author} {\bibfnamefont
  {R.}~\bibnamefont {Thomale}}, \bibinfo {author} {\bibfnamefont
  {C.}~\bibnamefont {Franchini}}, \bibinfo {author} {\bibfnamefont
  {S.}~\bibnamefont {Picozzi}}, \bibinfo {author} {\bibfnamefont
  {J.}~\bibnamefont {Fujii}}, \bibinfo {author} {\bibfnamefont {V.~N.}\
  \bibnamefont {Strocov}}, \bibinfo {author} {\bibfnamefont {G.}~\bibnamefont
  {Sangiovanni}}, \bibinfo {author} {\bibfnamefont {I.}~\bibnamefont
  {Vobornik}}, \bibinfo {author} {\bibfnamefont {R.~J.}\ \bibnamefont {Cava}},\
  and\ \bibinfo {author} {\bibfnamefont {G.}~\bibnamefont {Panaccione}},\
  }\bibfield  {title} {\bibinfo {title} {Three-{Dimensional} {Electronic}
  {Structure} of the {Type}-{II} {Weyl} {Semimetal} ${WTe}_{2}$},\ }\href
  {https://doi.org/10.1103/PhysRevLett.119.026403} {\bibfield  {journal}
  {\bibinfo  {journal} {Physical Review Letters}\ }\textbf {\bibinfo {volume}
  {119}},\ \bibinfo {pages} {026403} (\bibinfo {year} {2017})}\BibitemShut
  {NoStop}%
\bibitem [{\citenamefont {Wu}\ \emph {et~al.}(2019)\citenamefont {Wu},
  \citenamefont {Lovorn}, \citenamefont {Tutuc}, \citenamefont {Martin},\ and\
  \citenamefont {MacDonald}}]{Wu2019}%
  \BibitemOpen
  \bibfield  {author} {\bibinfo {author} {\bibfnamefont {F.}~\bibnamefont
  {Wu}}, \bibinfo {author} {\bibfnamefont {T.}~\bibnamefont {Lovorn}}, \bibinfo
  {author} {\bibfnamefont {E.}~\bibnamefont {Tutuc}}, \bibinfo {author}
  {\bibfnamefont {I.}~\bibnamefont {Martin}},\ and\ \bibinfo {author}
  {\bibfnamefont {A.~H.}\ \bibnamefont {MacDonald}},\ }\bibfield  {title}
  {\bibinfo {title} {Topological {Insulators} in {Twisted} {Transition} {Metal}
  {Dichalcogenide} {Homobilayers}},\ }\href
  {https://doi.org/10.1103/PhysRevLett.122.086402} {\bibfield  {journal}
  {\bibinfo  {journal} {Physical Review Letters}\ }\textbf {\bibinfo {volume}
  {122}},\ \bibinfo {pages} {086402} (\bibinfo {year} {2019})}\BibitemShut
  {NoStop}%
\bibitem [{\citenamefont {Wu}\ \emph {et~al.}(2018{\natexlab{a}})\citenamefont
  {Wu}, \citenamefont {Fatemi}, \citenamefont {Gibson}, \citenamefont
  {Watanabe}, \citenamefont {Taniguchi}, \citenamefont {Cava},\ and\
  \citenamefont {Jarillo-Herrero}}]{Wu2018}%
  \BibitemOpen
  \bibfield  {author} {\bibinfo {author} {\bibfnamefont {S.}~\bibnamefont
  {Wu}}, \bibinfo {author} {\bibfnamefont {V.}~\bibnamefont {Fatemi}}, \bibinfo
  {author} {\bibfnamefont {Q.~D.}\ \bibnamefont {Gibson}}, \bibinfo {author}
  {\bibfnamefont {K.}~\bibnamefont {Watanabe}}, \bibinfo {author}
  {\bibfnamefont {T.}~\bibnamefont {Taniguchi}}, \bibinfo {author}
  {\bibfnamefont {R.~J.}\ \bibnamefont {Cava}},\ and\ \bibinfo {author}
  {\bibfnamefont {P.}~\bibnamefont {Jarillo-Herrero}},\ }\bibfield  {title}
  {\bibinfo {title} {Observation of the quantum spin {Hall} effect up to 100
  kelvin in a monolayer crystal},\ }\href
  {https://doi.org/10.1126/science.aan6003} {\bibfield  {journal} {\bibinfo
  {journal} {Science}\ }\textbf {\bibinfo {volume} {359}},\ \bibinfo {pages}
  {76} (\bibinfo {year} {2018}{\natexlab{a}})}\BibitemShut {NoStop}%
\bibitem [{\citenamefont {Costanzo}\ \emph {et~al.}(2016)\citenamefont
  {Costanzo}, \citenamefont {Jo}, \citenamefont {Berger},\ and\ \citenamefont
  {Morpurgo}}]{Costanzo2016}%
  \BibitemOpen
  \bibfield  {author} {\bibinfo {author} {\bibfnamefont {D.}~\bibnamefont
  {Costanzo}}, \bibinfo {author} {\bibfnamefont {S.}~\bibnamefont {Jo}},
  \bibinfo {author} {\bibfnamefont {H.}~\bibnamefont {Berger}},\ and\ \bibinfo
  {author} {\bibfnamefont {A.~F.}\ \bibnamefont {Morpurgo}},\ }\bibfield
  {title} {\bibinfo {title} {Gate-induced superconductivity in atomically thin
  {MoS2} crystals},\ }\href {https://doi.org/10.1038/nnano.2015.314} {\bibfield
   {journal} {\bibinfo  {journal} {Nature Nanotechnology}\ }\textbf {\bibinfo
  {volume} {11}},\ \bibinfo {pages} {339} (\bibinfo {year} {2016})}\BibitemShut
  {NoStop}%
\bibitem [{\citenamefont {Fang}\ \emph {et~al.}(2018)\citenamefont {Fang},
  \citenamefont {Pan}, \citenamefont {He}, \citenamefont {Luo}, \citenamefont
  {Wang}, \citenamefont {Che}, \citenamefont {Bu}, \citenamefont {Zhao},
  \citenamefont {Liu}, \citenamefont {Mu}, \citenamefont {Zhang}, \citenamefont
  {Lin},\ and\ \citenamefont {Huang}}]{Fang2018}%
  \BibitemOpen
  \bibfield  {author} {\bibinfo {author} {\bibfnamefont {Y.}~\bibnamefont
  {Fang}}, \bibinfo {author} {\bibfnamefont {J.}~\bibnamefont {Pan}}, \bibinfo
  {author} {\bibfnamefont {J.}~\bibnamefont {He}}, \bibinfo {author}
  {\bibfnamefont {R.}~\bibnamefont {Luo}}, \bibinfo {author} {\bibfnamefont
  {D.}~\bibnamefont {Wang}}, \bibinfo {author} {\bibfnamefont {X.}~\bibnamefont
  {Che}}, \bibinfo {author} {\bibfnamefont {K.}~\bibnamefont {Bu}}, \bibinfo
  {author} {\bibfnamefont {W.}~\bibnamefont {Zhao}}, \bibinfo {author}
  {\bibfnamefont {P.}~\bibnamefont {Liu}}, \bibinfo {author} {\bibfnamefont
  {G.}~\bibnamefont {Mu}}, \bibinfo {author} {\bibfnamefont {H.}~\bibnamefont
  {Zhang}}, \bibinfo {author} {\bibfnamefont {T.}~\bibnamefont {Lin}},\ and\
  \bibinfo {author} {\bibfnamefont {F.}~\bibnamefont {Huang}},\ }\bibfield
  {title} {\bibinfo {title} {Structure re-determination and superconductivity
  observation of bulk 1t mos2},\ }\href
  {https://doi.org/10.1002/anie.201710512} {\bibfield  {journal} {\bibinfo
  {journal} {Angewandte Chemie International Edition}\ }\textbf {\bibinfo
  {volume} {57}},\ \bibinfo {pages} {1232} (\bibinfo {year}
  {2018})}\BibitemShut {NoStop}%
\bibitem [{\citenamefont {Bruyer}\ \emph {et~al.}(2016)\citenamefont {Bruyer},
  \citenamefont {Di~Sante}, \citenamefont {Barone}, \citenamefont {Stroppa},
  \citenamefont {Whangbo},\ and\ \citenamefont {Picozzi}}]{Bruyer2016}%
  \BibitemOpen
  \bibfield  {author} {\bibinfo {author} {\bibfnamefont {E.}~\bibnamefont
  {Bruyer}}, \bibinfo {author} {\bibfnamefont {D.}~\bibnamefont {Di~Sante}},
  \bibinfo {author} {\bibfnamefont {P.}~\bibnamefont {Barone}}, \bibinfo
  {author} {\bibfnamefont {A.}~\bibnamefont {Stroppa}}, \bibinfo {author}
  {\bibfnamefont {M.-H.}\ \bibnamefont {Whangbo}},\ and\ \bibinfo {author}
  {\bibfnamefont {S.}~\bibnamefont {Picozzi}},\ }\bibfield  {title} {\bibinfo
  {title} {Possibility of combining ferroelectricity and {Rashba}-like spin
  splitting in monolayers of the {1T}-type transition-metal dichalcogenides
  {M}{X}$_{2}$({M}={Mo},{W};{X}={{S}},{{Se}},{{Te}})},\ }\href
  {https://doi.org/10.1103/PhysRevB.94.195402} {\bibfield  {journal} {\bibinfo
  {journal} {Physical Review B}\ }\textbf {\bibinfo {volume} {94}},\ \bibinfo
  {pages} {195402} (\bibinfo {year} {2016})}\BibitemShut {NoStop}%
\bibitem [{\citenamefont {Strachan}\ \emph {et~al.}(2021)\citenamefont
  {Strachan}, \citenamefont {Masters},\ and\ \citenamefont
  {Maschmeyer}}]{Strachan2021}%
  \BibitemOpen
  \bibfield  {author} {\bibinfo {author} {\bibfnamefont {J.}~\bibnamefont
  {Strachan}}, \bibinfo {author} {\bibfnamefont {A.~F.}\ \bibnamefont
  {Masters}},\ and\ \bibinfo {author} {\bibfnamefont {T.}~\bibnamefont
  {Maschmeyer}},\ }\bibfield  {title} {\bibinfo {title} {{3R}-{MoS2} in
  {Review}: {History}, {Status}, and {Outlook}},\ }\href
  {https://doi.org/10.1021/acsaem.1c00638} {\bibfield  {journal} {\bibinfo
  {journal} {ACS Applied Energy Materials}\ }\textbf {\bibinfo {volume} {4}},\
  \bibinfo {pages} {7405} (\bibinfo {year} {2021})}\BibitemShut {NoStop}%
\bibitem [{\citenamefont {Lian}\ \emph {et~al.}(2020)\citenamefont {Lian},
  \citenamefont {Si},\ and\ \citenamefont {Duan}}]{Lian2021}%
  \BibitemOpen
  \bibfield  {author} {\bibinfo {author} {\bibfnamefont {C.-S.}\ \bibnamefont
  {Lian}}, \bibinfo {author} {\bibfnamefont {C.}~\bibnamefont {Si}},\ and\
  \bibinfo {author} {\bibfnamefont {W.}~\bibnamefont {Duan}},\ }\bibfield
  {title} {\bibinfo {title} {Anisotropic {Full}-{Gap} {Superconductivity} in
  {2M}-{WS2} {Topological} {Metal} with {Intrinsic} {Proximity} {Effect}},\
  }\href {https://doi.org/10.1021/acs.nanolett.0c04357} {\bibfield  {journal}
  {\bibinfo  {journal} {Nano Letters}\ }\textbf {\bibinfo {volume} {21}},\
  \bibinfo {pages} {709} (\bibinfo {year} {2020})}\BibitemShut {NoStop}%
\bibitem [{\citenamefont {He}\ \emph {et~al.}(2018)\citenamefont {He},
  \citenamefont {Zhou}, \citenamefont {He}, \citenamefont {Yuan}, \citenamefont
  {Zhang},\ and\ \citenamefont {Law}}]{He2018}%
  \BibitemOpen
  \bibfield  {author} {\bibinfo {author} {\bibfnamefont {W.-Y.}\ \bibnamefont
  {He}}, \bibinfo {author} {\bibfnamefont {B.~T.}\ \bibnamefont {Zhou}},
  \bibinfo {author} {\bibfnamefont {J.~J.}\ \bibnamefont {He}}, \bibinfo
  {author} {\bibfnamefont {N.~F.~Q.}\ \bibnamefont {Yuan}}, \bibinfo {author}
  {\bibfnamefont {T.}~\bibnamefont {Zhang}},\ and\ \bibinfo {author}
  {\bibfnamefont {K.~T.}\ \bibnamefont {Law}},\ }\bibfield  {title} {\bibinfo
  {title} {Magnetic field driven nodal topological superconductivity in
  monolayer transition metal dichalcogenides},\ }\href
  {https://doi.org/10.1038/s42005-018-0041-4} {\bibfield  {journal} {\bibinfo
  {journal} {Communications Physics}\ }\textbf {\bibinfo {volume} {1}},\
  \bibinfo {pages} {1} (\bibinfo {year} {2018})}\BibitemShut {NoStop}%
\bibitem [{\citenamefont {Zhao}\ \emph {et~al.}(2020)\citenamefont {Zhao},
  \citenamefont {Liu}, \citenamefont {Li}, \citenamefont {Ren}, \citenamefont
  {Chen}, \citenamefont {Yu},\ and\ \citenamefont {Zhang}}]{Zhao2020}%
  \BibitemOpen
  \bibfield  {author} {\bibinfo {author} {\bibfnamefont {C.-X.}\ \bibnamefont
  {Zhao}}, \bibinfo {author} {\bibfnamefont {J.-N.}\ \bibnamefont {Liu}},
  \bibinfo {author} {\bibfnamefont {B.-Q.}\ \bibnamefont {Li}}, \bibinfo
  {author} {\bibfnamefont {D.}~\bibnamefont {Ren}}, \bibinfo {author}
  {\bibfnamefont {X.}~\bibnamefont {Chen}}, \bibinfo {author} {\bibfnamefont
  {J.}~\bibnamefont {Yu}},\ and\ \bibinfo {author} {\bibfnamefont
  {Q.}~\bibnamefont {Zhang}},\ }\bibfield  {title} {\bibinfo {title}
  {Multiscale {Construction} of {Bifunctional} {Electrocatalysts} for
  {Long}-{Lifespan} {Rechargeable} {Zinc}–{Air} {Batteries}},\ }\href
  {https://doi.org/10.1002/adfm.202003619} {\bibfield  {journal} {\bibinfo
  {journal} {Advanced Functional Materials}\ }\textbf {\bibinfo {volume}
  {30}},\ \bibinfo {pages} {2003619} (\bibinfo {year} {2020})}\BibitemShut
  {NoStop}%
\bibitem [{\citenamefont {Kusmartseva}\ \emph {et~al.}(2009)\citenamefont
  {Kusmartseva}, \citenamefont {Sipos}, \citenamefont {Berger}, \citenamefont
  {Forro},\ and\ \citenamefont {Tutis}}]{Kusmartseva2009}%
  \BibitemOpen
  \bibfield  {author} {\bibinfo {author} {\bibfnamefont {A.~F.}\ \bibnamefont
  {Kusmartseva}}, \bibinfo {author} {\bibfnamefont {B.}~\bibnamefont {Sipos}},
  \bibinfo {author} {\bibfnamefont {H.}~\bibnamefont {Berger}}, \bibinfo
  {author} {\bibfnamefont {L.}~\bibnamefont {Forro}},\ and\ \bibinfo {author}
  {\bibfnamefont {E.}~\bibnamefont {Tutis}},\ }\bibfield  {title} {\bibinfo
  {title} {Pressure {Induced} {Superconductivity} in {Pristine}
  ${1T}{-}{TiSe}_{2}$},\ }\href
  {https://doi.org/10.1103/PhysRevLett.103.236401} {\bibfield  {journal}
  {\bibinfo  {journal} {Physical Review Letters}\ }\textbf {\bibinfo {volume}
  {103}},\ \bibinfo {pages} {236401} (\bibinfo {year} {2009})}\BibitemShut
  {NoStop}%
\bibitem [{\citenamefont {Xiao}\ \emph {et~al.}(2017)\citenamefont {Xiao},
  \citenamefont {Gong}, \citenamefont {Wu}, \citenamefont {Lu}, \citenamefont
  {Wei}, \citenamefont {Li}, \citenamefont {Lv}, \citenamefont {Luo},
  \citenamefont {Tong}, \citenamefont {Zhu},\ and\ \citenamefont
  {Sun}}]{Xiao2017}%
  \BibitemOpen
  \bibfield  {author} {\bibinfo {author} {\bibfnamefont {R.~C.}\ \bibnamefont
  {Xiao}}, \bibinfo {author} {\bibfnamefont {P.~L.}\ \bibnamefont {Gong}},
  \bibinfo {author} {\bibfnamefont {Q.~S.}\ \bibnamefont {Wu}}, \bibinfo
  {author} {\bibfnamefont {W.~J.}\ \bibnamefont {Lu}}, \bibinfo {author}
  {\bibfnamefont {M.~J.}\ \bibnamefont {Wei}}, \bibinfo {author} {\bibfnamefont
  {J.~Y.}\ \bibnamefont {Li}}, \bibinfo {author} {\bibfnamefont {H.~Y.}\
  \bibnamefont {Lv}}, \bibinfo {author} {\bibfnamefont {X.}~\bibnamefont
  {Luo}}, \bibinfo {author} {\bibfnamefont {P.}~\bibnamefont {Tong}}, \bibinfo
  {author} {\bibfnamefont {X.~B.}\ \bibnamefont {Zhu}},\ and\ \bibinfo {author}
  {\bibfnamefont {Y.~P.}\ \bibnamefont {Sun}},\ }\bibfield  {title} {\bibinfo
  {title} {Manipulation of type-{I} and type-{II} {Dirac} points in
  ${PdTe}_{2}$ superconductor by external pressure},\ }\href
  {https://doi.org/10.1103/PhysRevB.96.075101} {\bibfield  {journal} {\bibinfo
  {journal} {Physical Review B}\ }\textbf {\bibinfo {volume} {96}},\ \bibinfo
  {pages} {075101} (\bibinfo {year} {2017})}\BibitemShut {NoStop}%
\bibitem [{\citenamefont {He}\ \emph {et~al.}(2014)\citenamefont {He},
  \citenamefont {Liu}, \citenamefont {He}, \citenamefont {Lai}, \citenamefont
  {He}, \citenamefont {Wang}, \citenamefont {Law}, \citenamefont {Lortz},
  \citenamefont {Wang},\ and\ \citenamefont {Sou}}]{He2014}%
  \BibitemOpen
  \bibfield  {author} {\bibinfo {author} {\bibfnamefont {Q.~L.}\ \bibnamefont
  {He}}, \bibinfo {author} {\bibfnamefont {H.}~\bibnamefont {Liu}}, \bibinfo
  {author} {\bibfnamefont {M.}~\bibnamefont {He}}, \bibinfo {author}
  {\bibfnamefont {Y.~H.}\ \bibnamefont {Lai}}, \bibinfo {author} {\bibfnamefont
  {H.}~\bibnamefont {He}}, \bibinfo {author} {\bibfnamefont {G.}~\bibnamefont
  {Wang}}, \bibinfo {author} {\bibfnamefont {K.~T.}\ \bibnamefont {Law}},
  \bibinfo {author} {\bibfnamefont {R.}~\bibnamefont {Lortz}}, \bibinfo
  {author} {\bibfnamefont {J.}~\bibnamefont {Wang}},\ and\ \bibinfo {author}
  {\bibfnamefont {I.~K.}\ \bibnamefont {Sou}},\ }\bibfield  {title} {\bibinfo
  {title} {Two-dimensional superconductivity at the interface of a
  {Bi2Te3}/{FeTe} heterostructure},\ }\href
  {https://doi.org/10.1038/ncomms5247} {\bibfield  {journal} {\bibinfo
  {journal} {Nature Communications}\ }\textbf {\bibinfo {volume} {5}},\
  \bibinfo {pages} {4247} (\bibinfo {year} {2014})}\BibitemShut {NoStop}%
\bibitem [{\citenamefont {Ermolaev}\ \emph {et~al.}(2022)\citenamefont
  {Ermolaev}, \citenamefont {Voronin}, \citenamefont {Baranov}, \citenamefont
  {Kravets}, \citenamefont {Tselikov}, \citenamefont {Stebunov}, \citenamefont
  {Yakubovsky}, \citenamefont {Novikov}, \citenamefont {Vyshnevyy},
  \citenamefont {Mazitov}, \citenamefont {Kruglov}, \citenamefont {Zhukov},
  \citenamefont {Romanov}, \citenamefont {Markeev}, \citenamefont {Arsenin},
  \citenamefont {Novoselov}, \citenamefont {Grigorenko},\ and\ \citenamefont
  {Volkov}}]{Ermolaev2022}%
  \BibitemOpen
  \bibfield  {author} {\bibinfo {author} {\bibfnamefont {G.}~\bibnamefont
  {Ermolaev}}, \bibinfo {author} {\bibfnamefont {K.}~\bibnamefont {Voronin}},
  \bibinfo {author} {\bibfnamefont {D.~G.}\ \bibnamefont {Baranov}}, \bibinfo
  {author} {\bibfnamefont {V.}~\bibnamefont {Kravets}}, \bibinfo {author}
  {\bibfnamefont {G.}~\bibnamefont {Tselikov}}, \bibinfo {author}
  {\bibfnamefont {Y.}~\bibnamefont {Stebunov}}, \bibinfo {author}
  {\bibfnamefont {D.}~\bibnamefont {Yakubovsky}}, \bibinfo {author}
  {\bibfnamefont {S.}~\bibnamefont {Novikov}}, \bibinfo {author} {\bibfnamefont
  {A.}~\bibnamefont {Vyshnevyy}}, \bibinfo {author} {\bibfnamefont
  {A.}~\bibnamefont {Mazitov}}, \bibinfo {author} {\bibfnamefont
  {I.}~\bibnamefont {Kruglov}}, \bibinfo {author} {\bibfnamefont
  {S.}~\bibnamefont {Zhukov}}, \bibinfo {author} {\bibfnamefont
  {R.}~\bibnamefont {Romanov}}, \bibinfo {author} {\bibfnamefont {A.~M.}\
  \bibnamefont {Markeev}}, \bibinfo {author} {\bibfnamefont {A.}~\bibnamefont
  {Arsenin}}, \bibinfo {author} {\bibfnamefont {K.~S.}\ \bibnamefont
  {Novoselov}}, \bibinfo {author} {\bibfnamefont {A.~N.}\ \bibnamefont
  {Grigorenko}},\ and\ \bibinfo {author} {\bibfnamefont {V.}~\bibnamefont
  {Volkov}},\ }\bibfield  {title} {\bibinfo {title} {Topological phase
  singularities in atomically thin high-refractive-index materials},\ }\href
  {https://doi.org/10.1038/s41467-022-29716-4} {\bibfield  {journal} {\bibinfo
  {journal} {Nature Communications}\ }\textbf {\bibinfo {volume} {13}},\
  \bibinfo {pages} {2049} (\bibinfo {year} {2022})}\BibitemShut {NoStop}%
\bibitem [{\citenamefont {Yan}\ \emph {et~al.}(2017)\citenamefont {Yan},
  \citenamefont {Huang}, \citenamefont {Zhang}, \citenamefont {Wang},
  \citenamefont {Yao}, \citenamefont {Deng}, \citenamefont {Wan}, \citenamefont
  {Zhang}, \citenamefont {Arita}, \citenamefont {Yang}, \citenamefont {Sun},
  \citenamefont {Yao}, \citenamefont {Wu}, \citenamefont {Fan}, \citenamefont
  {Duan},\ and\ \citenamefont {Zhou}}]{Yan2017}%
  \BibitemOpen
  \bibfield  {author} {\bibinfo {author} {\bibfnamefont {M.}~\bibnamefont
  {Yan}}, \bibinfo {author} {\bibfnamefont {H.}~\bibnamefont {Huang}}, \bibinfo
  {author} {\bibfnamefont {K.}~\bibnamefont {Zhang}}, \bibinfo {author}
  {\bibfnamefont {E.}~\bibnamefont {Wang}}, \bibinfo {author} {\bibfnamefont
  {W.}~\bibnamefont {Yao}}, \bibinfo {author} {\bibfnamefont {K.}~\bibnamefont
  {Deng}}, \bibinfo {author} {\bibfnamefont {G.}~\bibnamefont {Wan}}, \bibinfo
  {author} {\bibfnamefont {H.}~\bibnamefont {Zhang}}, \bibinfo {author}
  {\bibfnamefont {M.}~\bibnamefont {Arita}}, \bibinfo {author} {\bibfnamefont
  {H.}~\bibnamefont {Yang}}, \bibinfo {author} {\bibfnamefont {Z.}~\bibnamefont
  {Sun}}, \bibinfo {author} {\bibfnamefont {H.}~\bibnamefont {Yao}}, \bibinfo
  {author} {\bibfnamefont {Y.}~\bibnamefont {Wu}}, \bibinfo {author}
  {\bibfnamefont {S.}~\bibnamefont {Fan}}, \bibinfo {author} {\bibfnamefont
  {W.}~\bibnamefont {Duan}},\ and\ \bibinfo {author} {\bibfnamefont
  {S.}~\bibnamefont {Zhou}},\ }\bibfield  {title} {\bibinfo {title}
  {Lorentz-violating type-{II} {Dirac} fermions in transition metal
  dichalcogenide {PtTe2}},\ }\bibfield  {journal} {\bibinfo  {journal} {Nature
  Communications}\ }\textbf {\bibinfo {volume} {8}},\ \href
  {https://doi.org/10.1038/s41467-017-00280-6} {10.1038/s41467-017-00280-6}
  (\bibinfo {year} {2017})\BibitemShut {NoStop}%
\bibitem [{\citenamefont {Tanisha}\ \emph {et~al.}(2024)\citenamefont
  {Tanisha}, \citenamefont {Hossain}, \citenamefont {Hiramony}, \citenamefont
  {Rasul}, \citenamefont {Hasan},\ and\ \citenamefont {Khosru}}]{Tanisha2024}%
  \BibitemOpen
  \bibfield  {author} {\bibinfo {author} {\bibfnamefont {T.~T.}\ \bibnamefont
  {Tanisha}}, \bibinfo {author} {\bibfnamefont {M.~S.}\ \bibnamefont
  {Hossain}}, \bibinfo {author} {\bibfnamefont {N.~T.}\ \bibnamefont
  {Hiramony}}, \bibinfo {author} {\bibfnamefont {A.}~\bibnamefont {Rasul}},
  \bibinfo {author} {\bibfnamefont {M.~Z.}\ \bibnamefont {Hasan}},\ and\
  \bibinfo {author} {\bibfnamefont {Q.~D.~M.}\ \bibnamefont {Khosru}},\ }\href
  {https://doi.org/10.48550/arXiv.2401.13124} {\emph {\bibinfo {title} {Tunable
  {Topological} {Phase} {Transitions} in a {Piezoelectric} {Janus}
  {Monolayer}}}},\ \bibinfo {type} {Tech. Rep.}\ (\bibinfo  {institution}
  {arXiv.2401.13124},\ \bibinfo {year} {2024})\BibitemShut {NoStop}%
\bibitem [{\citenamefont {Sato}\ and\ \citenamefont {Ando}(2017)}]{Sato2017}%
  \BibitemOpen
  \bibfield  {author} {\bibinfo {author} {\bibfnamefont {M.}~\bibnamefont
  {Sato}}\ and\ \bibinfo {author} {\bibfnamefont {Y.}~\bibnamefont {Ando}},\
  }\bibfield  {title} {\bibinfo {title} {Topological superconductors: a
  review},\ }\href {https://doi.org/10.1088/1361-6633/aa6ac7} {\bibfield
  {journal} {\bibinfo  {journal} {Reports on Progress in Physics}\ }\textbf
  {\bibinfo {volume} {80}},\ \bibinfo {pages} {076501} (\bibinfo {year}
  {2017})}\BibitemShut {NoStop}%
\bibitem [{\citenamefont {Gu}\ \emph {et~al.}(2020)\citenamefont {Gu},
  \citenamefont {Luo}, \citenamefont {Ge},\ and\ \citenamefont
  {Wang}}]{Gu2020}%
  \BibitemOpen
  \bibfield  {author} {\bibinfo {author} {\bibfnamefont {K.-Y.}\ \bibnamefont
  {Gu}}, \bibinfo {author} {\bibfnamefont {T.-C.}\ \bibnamefont {Luo}},
  \bibinfo {author} {\bibfnamefont {J.}~\bibnamefont {Ge}},\ and\ \bibinfo
  {author} {\bibfnamefont {J.}~\bibnamefont {Wang}},\ }\bibfield  {title}
  {\bibinfo {title} {Superconductivity in topological materials},\ }\href
  {https://doi.org/10.7498/aps.69.20191627} {\bibfield  {journal} {\bibinfo
  {journal} {Acta Physica Sinica}\ }\textbf {\bibinfo {volume} {69}},\ \bibinfo
  {pages} {020301} (\bibinfo {year} {2020})}\BibitemShut {NoStop}%
\bibitem [{\citenamefont {Zhang}\ \emph {et~al.}(2020)\citenamefont {Zhang},
  \citenamefont {Rui}, \citenamefont {Calzona}, \citenamefont {Choi},
  \citenamefont {Schnyder},\ and\ \citenamefont {Trauzettel}}]{Zhang2020}%
  \BibitemOpen
  \bibfield  {author} {\bibinfo {author} {\bibfnamefont {S.-B.}\ \bibnamefont
  {Zhang}}, \bibinfo {author} {\bibfnamefont {W.~B.}\ \bibnamefont {Rui}},
  \bibinfo {author} {\bibfnamefont {A.}~\bibnamefont {Calzona}}, \bibinfo
  {author} {\bibfnamefont {S.-J.}\ \bibnamefont {Choi}}, \bibinfo {author}
  {\bibfnamefont {A.~P.}\ \bibnamefont {Schnyder}},\ and\ \bibinfo {author}
  {\bibfnamefont {B.}~\bibnamefont {Trauzettel}},\ }\bibfield  {title}
  {\bibinfo {title} {Topological and holonomic quantum computation based on
  second-order topological superconductors},\ }\href
  {https://doi.org/10.1103/PhysRevResearch.2.043025} {\bibfield  {journal}
  {\bibinfo  {journal} {Physical Review Research}\ }\textbf {\bibinfo {volume}
  {2}},\ \bibinfo {pages} {043025} (\bibinfo {year} {2020})}\BibitemShut
  {NoStop}%
\bibitem [{\citenamefont {Kitaev}(2003)}]{Kitaev2003}%
  \BibitemOpen
  \bibfield  {author} {\bibinfo {author} {\bibfnamefont {A.~Y.}\ \bibnamefont
  {Kitaev}},\ }\bibfield  {title} {\bibinfo {title} {Fault-tolerant quantum
  computation by anyons},\ }\href
  {https://doi.org/10.1016/S0003-4916(02)00018-0} {\bibfield  {journal}
  {\bibinfo  {journal} {Annals of Physics}\ }\textbf {\bibinfo {volume}
  {303}},\ \bibinfo {pages} {2} (\bibinfo {year} {2003})}\BibitemShut {NoStop}%
\bibitem [{\citenamefont {Lahtinen}\ and\ \citenamefont
  {Pachos}(2017)}]{Lahtinen2017}%
  \BibitemOpen
  \bibfield  {author} {\bibinfo {author} {\bibfnamefont {V.}~\bibnamefont
  {Lahtinen}}\ and\ \bibinfo {author} {\bibfnamefont {J.}~\bibnamefont
  {Pachos}},\ }\bibfield  {title} {\bibinfo {title} {A {Short} {Introduction}
  to {Topological} {Quantum} {Computation}},\ }\href
  {https://doi.org/10.21468/SciPostPhys.3.3.021} {\bibfield  {journal}
  {\bibinfo  {journal} {SciPost Physics}\ }\textbf {\bibinfo {volume} {3}},\
  \bibinfo {pages} {021} (\bibinfo {year} {2017})}\BibitemShut {NoStop}%
\bibitem [{\citenamefont {Nayak}\ \emph {et~al.}(2008)\citenamefont {Nayak},
  \citenamefont {Simon}, \citenamefont {Stern}, \citenamefont {Freedman},\ and\
  \citenamefont {Das~Sarma}}]{Nayak2008}%
  \BibitemOpen
  \bibfield  {author} {\bibinfo {author} {\bibfnamefont {C.}~\bibnamefont
  {Nayak}}, \bibinfo {author} {\bibfnamefont {S.~H.}\ \bibnamefont {Simon}},
  \bibinfo {author} {\bibfnamefont {A.}~\bibnamefont {Stern}}, \bibinfo
  {author} {\bibfnamefont {M.}~\bibnamefont {Freedman}},\ and\ \bibinfo
  {author} {\bibfnamefont {S.}~\bibnamefont {Das~Sarma}},\ }\bibfield  {title}
  {\bibinfo {title} {Non-{Abelian} anyons and topological quantum
  computation},\ }\href {https://doi.org/10.1103/RevModPhys.80.1083} {\bibfield
   {journal} {\bibinfo  {journal} {Reviews of Modern Physics}\ }\textbf
  {\bibinfo {volume} {80}},\ \bibinfo {pages} {1083} (\bibinfo {year}
  {2008})}\BibitemShut {NoStop}%
\bibitem [{\citenamefont {Sarma}\ \emph {et~al.}(2015)\citenamefont {Sarma},
  \citenamefont {Freedman},\ and\ \citenamefont {Nayak}}]{Sarma2015}%
  \BibitemOpen
  \bibfield  {author} {\bibinfo {author} {\bibfnamefont {S.~D.}\ \bibnamefont
  {Sarma}}, \bibinfo {author} {\bibfnamefont {M.}~\bibnamefont {Freedman}},\
  and\ \bibinfo {author} {\bibfnamefont {C.}~\bibnamefont {Nayak}},\ }\bibfield
   {title} {\bibinfo {title} {Majorana zero modes and topological quantum
  computation},\ }\href {https://doi.org/10.1038/npjqi.2015.1} {\bibfield
  {journal} {\bibinfo  {journal} {npj Quantum Information}\ }\textbf {\bibinfo
  {volume} {1}},\ \bibinfo {pages} {1} (\bibinfo {year} {2015})}\BibitemShut
  {NoStop}%
\bibitem [{\citenamefont {Fidkowski}(2010)}]{Fidkowski2010}%
  \BibitemOpen
  \bibfield  {author} {\bibinfo {author} {\bibfnamefont {L.}~\bibnamefont
  {Fidkowski}},\ }\bibfield  {title} {\bibinfo {title} {Entanglement {Spectrum}
  of {Topological} {Insulators} and {Superconductors}},\ }\href
  {https://doi.org/10.1103/PhysRevLett.104.130502} {\bibfield  {journal}
  {\bibinfo  {journal} {Physical Review Letters}\ }\textbf {\bibinfo {volume}
  {104}},\ \bibinfo {pages} {130502} (\bibinfo {year} {2010})}\BibitemShut
  {NoStop}%
\bibitem [{\citenamefont {Qi}\ and\ \citenamefont {Zhang}(2011)}]{Qi2011}%
  \BibitemOpen
  \bibfield  {author} {\bibinfo {author} {\bibfnamefont {X.-L.}\ \bibnamefont
  {Qi}}\ and\ \bibinfo {author} {\bibfnamefont {S.-C.}\ \bibnamefont {Zhang}},\
  }\bibfield  {title} {\bibinfo {title} {Topological insulators and
  superconductors},\ }\href {https://doi.org/10.1103/RevModPhys.83.1057}
  {\bibfield  {journal} {\bibinfo  {journal} {Reviews of Modern Physics}\
  }\textbf {\bibinfo {volume} {83}},\ \bibinfo {pages} {1057} (\bibinfo {year}
  {2011})}\BibitemShut {NoStop}%
\bibitem [{\citenamefont {Patra}\ \emph {et~al.}(2023)\citenamefont {Patra},
  \citenamefont {Agarwal}, \citenamefont {Arushi}, \citenamefont {Manna},
  \citenamefont {Bhatt}, \citenamefont {Singh},\ and\ \citenamefont
  {Singh}}]{Patra2023}%
  \BibitemOpen
  \bibfield  {author} {\bibinfo {author} {\bibfnamefont {C.}~\bibnamefont
  {Patra}}, \bibinfo {author} {\bibfnamefont {T.}~\bibnamefont {Agarwal}},
  \bibinfo {author} {\bibnamefont {Arushi}}, \bibinfo {author} {\bibfnamefont
  {P.}~\bibnamefont {Manna}}, \bibinfo {author} {\bibfnamefont
  {N.}~\bibnamefont {Bhatt}}, \bibinfo {author} {\bibfnamefont {R.~S.}\
  \bibnamefont {Singh}},\ and\ \bibinfo {author} {\bibfnamefont {R.~P.}\
  \bibnamefont {Singh}},\ }\href {https://doi.org/10.48550/arXiv.2311.01019}
  {\emph {\bibinfo {title} {Superconducting {Properties} of {Topological}
  {Semimetal} 1{T}-{RhSeTe}}}},\ \bibinfo {type} {Tech. Rep.}\ (\bibinfo
  {institution} {arXiv:2311.01019},\ \bibinfo {year} {2023})\BibitemShut
  {NoStop}%
\bibitem [{\citenamefont {Geller}(1955)}]{Geller1955}%
  \BibitemOpen
  \bibfield  {author} {\bibinfo {author} {\bibfnamefont {S.}~\bibnamefont
  {Geller}},\ }\bibfield  {title} {\bibinfo {title} {The {Crystal} {Structures}
  of {RhTe} and {RhTe2}},\ }\href {https://doi.org/10.1021/ja01614a091}
  {\bibfield  {journal} {\bibinfo  {journal} {Journal of the American Chemical
  Society}\ }\textbf {\bibinfo {volume} {77}},\ \bibinfo {pages} {2641}
  (\bibinfo {year} {1955})}\BibitemShut {NoStop}%
\bibitem [{\citenamefont {Lurgo}\ \emph {et~al.}(2022)\citenamefont {Lurgo},
  \citenamefont {Pomiro}, \citenamefont {Carbonio},\ and\ \citenamefont
  {Sanchez}}]{Lurgo2022}%
  \BibitemOpen
  \bibfield  {author} {\bibinfo {author} {\bibfnamefont {F.~E.}\ \bibnamefont
  {Lurgo}}, \bibinfo {author} {\bibfnamefont {F.}~\bibnamefont {Pomiro}},
  \bibinfo {author} {\bibfnamefont {R.~E.}\ \bibnamefont {Carbonio}},\ and\
  \bibinfo {author} {\bibfnamefont {R.~D.}\ \bibnamefont {Sanchez}},\
  }\bibfield  {title} {\bibinfo {title} {Synthesis and structural, magnetic,
  electric, and thermoelectric characterization of layered
  {Rh}\_{1-x}{Ir}\_{x}{Te}\_{2}(0{$\le$}{x}{$\le$}1)},\ }\href
  {https://doi.org/10.1103/PhysRevB.105.104104} {\bibfield  {journal} {\bibinfo
   {journal} {Physical Review B}\ }\textbf {\bibinfo {volume} {105}},\ \bibinfo
  {pages} {104104} (\bibinfo {year} {2022})}\BibitemShut {NoStop}%
\bibitem [{\citenamefont {Kresse}\ and\ \citenamefont
  {Hafner}(1993)}]{Kresse1993}%
  \BibitemOpen
  \bibfield  {author} {\bibinfo {author} {\bibfnamefont {G.}~\bibnamefont
  {Kresse}}\ and\ \bibinfo {author} {\bibfnamefont {J.}~\bibnamefont
  {Hafner}},\ }\bibfield  {title} {\bibinfo {title} {Ab initio molecular
  dynamics for liquid metals},\ }\href
  {https://doi.org/10.1103/physrevb.47.558} {\bibfield  {journal} {\bibinfo
  {journal} {Physical Review. B, Condensed Matter}\ }\textbf {\bibinfo {volume}
  {47}},\ \bibinfo {pages} {558} (\bibinfo {year} {1993})}\BibitemShut
  {NoStop}%
\bibitem [{\citenamefont {Perdew}\ \emph {et~al.}(1996)\citenamefont {Perdew},
  \citenamefont {Burke},\ and\ \citenamefont {Ernzerhof}}]{Perdew1996}%
  \BibitemOpen
  \bibfield  {author} {\bibinfo {author} {\bibfnamefont {J.~P.}\ \bibnamefont
  {Perdew}}, \bibinfo {author} {\bibfnamefont {K.}~\bibnamefont {Burke}},\ and\
  \bibinfo {author} {\bibfnamefont {M.}~\bibnamefont {Ernzerhof}},\ }\bibfield
  {title} {\bibinfo {title} {Generalized {Gradient} {Approximation} {Made}
  {Simple}},\ }\href {https://doi.org/10.1103/PhysRevLett.77.3865} {\bibfield
  {journal} {\bibinfo  {journal} {Physical Review Letters}\ }\textbf {\bibinfo
  {volume} {77}},\ \bibinfo {pages} {3865} (\bibinfo {year}
  {1996})}\BibitemShut {NoStop}%
\bibitem [{\citenamefont {Kresse}\ and\ \citenamefont
  {Joubert}(1999)}]{Kresse1999}%
  \BibitemOpen
  \bibfield  {author} {\bibinfo {author} {\bibfnamefont {G.}~\bibnamefont
  {Kresse}}\ and\ \bibinfo {author} {\bibfnamefont {D.}~\bibnamefont
  {Joubert}},\ }\bibfield  {title} {\bibinfo {title} {From ultrasoft
  pseudopotentials to the projector augmented-wave method},\ }\href
  {https://doi.org/10.1103/PhysRevB.59.1758} {\bibfield  {journal} {\bibinfo
  {journal} {Physical Review B}\ }\textbf {\bibinfo {volume} {59}},\ \bibinfo
  {pages} {1758} (\bibinfo {year} {1999})}\BibitemShut {NoStop}%
\bibitem [{\citenamefont {Klimeš}\ \emph {et~al.}(2009)\citenamefont
  {Klimeš}, \citenamefont {Bowler},\ and\ \citenamefont
  {Michaelides}}]{Klimes2009}%
  \BibitemOpen
  \bibfield  {author} {\bibinfo {author} {\bibfnamefont {J.}~\bibnamefont
  {Klimeš}}, \bibinfo {author} {\bibfnamefont {D.~R.}\ \bibnamefont
  {Bowler}},\ and\ \bibinfo {author} {\bibfnamefont {A.}~\bibnamefont
  {Michaelides}},\ }\bibfield  {title} {\bibinfo {title} {Chemical accuracy for
  the van der waals density functional},\ }\href
  {https://doi.org/10.1088/0953-8984/22/2/022201} {\bibfield  {journal}
  {\bibinfo  {journal} {Journal of Physics: Condensed Matter}\ }\textbf
  {\bibinfo {volume} {22}},\ \bibinfo {pages} {022201} (\bibinfo {year}
  {2009})}\BibitemShut {NoStop}%
\bibitem [{\citenamefont {Giannozzi}\ \emph {et~al.}(2017)\citenamefont
  {Giannozzi}, \citenamefont {Andreussi}, \citenamefont {Brumme}, \citenamefont
  {Bunau}, \citenamefont {Nardelli}, \citenamefont {Calandra}, \citenamefont
  {Car}, \citenamefont {Cavazzoni}, \citenamefont {Ceresoli}, \citenamefont
  {Cococcioni}, \citenamefont {Colonna}, \citenamefont {Carnimeo},
  \citenamefont {Corso}, \citenamefont {Gironcoli}, \citenamefont {Delugas},
  \citenamefont {DiStasio}, \citenamefont {Ferretti}, \citenamefont {Floris},
  \citenamefont {Fratesi}, \citenamefont {Fugallo}, \citenamefont {Gebauer},
  \citenamefont {Gerstmann}, \citenamefont {Giustino}, \citenamefont {Gorni},
  \citenamefont {Jia}, \citenamefont {Kawamura}, \citenamefont {Ko},
  \citenamefont {Kokalj}, \citenamefont {Küçükbenli}, \citenamefont
  {Lazzeri}, \citenamefont {Marsili}, \citenamefont {Marzari}, \citenamefont
  {Mauri}, \citenamefont {Nguyen}, \citenamefont {Nguyen}, \citenamefont
  {Otero-de-la Roza}, \citenamefont {Paulatto}, \citenamefont {Poncé},
  \citenamefont {Rocca}, \citenamefont {Sabatini}, \citenamefont {Santra},
  \citenamefont {Schlipf}, \citenamefont {Seitsonen}, \citenamefont {Smogunov},
  \citenamefont {Timrov}, \citenamefont {Thonhauser}, \citenamefont {Umari},
  \citenamefont {Vast}, \citenamefont {Wu},\ and\ \citenamefont
  {Baroni}}]{Giannozzi2017}%
  \BibitemOpen
  \bibfield  {author} {\bibinfo {author} {\bibfnamefont {P.}~\bibnamefont
  {Giannozzi}}, \bibinfo {author} {\bibfnamefont {O.}~\bibnamefont
  {Andreussi}}, \bibinfo {author} {\bibfnamefont {T.}~\bibnamefont {Brumme}},
  \bibinfo {author} {\bibfnamefont {O.}~\bibnamefont {Bunau}}, \bibinfo
  {author} {\bibfnamefont {M.~B.}\ \bibnamefont {Nardelli}}, \bibinfo {author}
  {\bibfnamefont {M.}~\bibnamefont {Calandra}}, \bibinfo {author}
  {\bibfnamefont {R.}~\bibnamefont {Car}}, \bibinfo {author} {\bibfnamefont
  {C.}~\bibnamefont {Cavazzoni}}, \bibinfo {author} {\bibfnamefont
  {D.}~\bibnamefont {Ceresoli}}, \bibinfo {author} {\bibfnamefont
  {M.}~\bibnamefont {Cococcioni}}, \bibinfo {author} {\bibfnamefont
  {N.}~\bibnamefont {Colonna}}, \bibinfo {author} {\bibfnamefont
  {I.}~\bibnamefont {Carnimeo}}, \bibinfo {author} {\bibfnamefont {A.~D.}\
  \bibnamefont {Corso}}, \bibinfo {author} {\bibfnamefont {S.~d.}\ \bibnamefont
  {Gironcoli}}, \bibinfo {author} {\bibfnamefont {P.}~\bibnamefont {Delugas}},
  \bibinfo {author} {\bibfnamefont {R.~A.}\ \bibnamefont {DiStasio}}, \bibinfo
  {author} {\bibfnamefont {A.}~\bibnamefont {Ferretti}}, \bibinfo {author}
  {\bibfnamefont {A.}~\bibnamefont {Floris}}, \bibinfo {author} {\bibfnamefont
  {G.}~\bibnamefont {Fratesi}}, \bibinfo {author} {\bibfnamefont
  {G.}~\bibnamefont {Fugallo}}, \bibinfo {author} {\bibfnamefont
  {R.}~\bibnamefont {Gebauer}}, \bibinfo {author} {\bibfnamefont
  {U.}~\bibnamefont {Gerstmann}}, \bibinfo {author} {\bibfnamefont
  {F.}~\bibnamefont {Giustino}}, \bibinfo {author} {\bibfnamefont
  {T.}~\bibnamefont {Gorni}}, \bibinfo {author} {\bibfnamefont
  {J.}~\bibnamefont {Jia}}, \bibinfo {author} {\bibfnamefont {M.}~\bibnamefont
  {Kawamura}}, \bibinfo {author} {\bibfnamefont {H.-Y.}\ \bibnamefont {Ko}},
  \bibinfo {author} {\bibfnamefont {A.}~\bibnamefont {Kokalj}}, \bibinfo
  {author} {\bibfnamefont {E.}~\bibnamefont {Küçükbenli}}, \bibinfo {author}
  {\bibfnamefont {M.}~\bibnamefont {Lazzeri}}, \bibinfo {author} {\bibfnamefont
  {M.}~\bibnamefont {Marsili}}, \bibinfo {author} {\bibfnamefont
  {N.}~\bibnamefont {Marzari}}, \bibinfo {author} {\bibfnamefont
  {F.}~\bibnamefont {Mauri}}, \bibinfo {author} {\bibfnamefont {N.~L.}\
  \bibnamefont {Nguyen}}, \bibinfo {author} {\bibfnamefont {H.-V.}\
  \bibnamefont {Nguyen}}, \bibinfo {author} {\bibfnamefont {A.}~\bibnamefont
  {Otero-de-la Roza}}, \bibinfo {author} {\bibfnamefont {L.}~\bibnamefont
  {Paulatto}}, \bibinfo {author} {\bibfnamefont {S.}~\bibnamefont {Poncé}},
  \bibinfo {author} {\bibfnamefont {D.}~\bibnamefont {Rocca}}, \bibinfo
  {author} {\bibfnamefont {R.}~\bibnamefont {Sabatini}}, \bibinfo {author}
  {\bibfnamefont {B.}~\bibnamefont {Santra}}, \bibinfo {author} {\bibfnamefont
  {M.}~\bibnamefont {Schlipf}}, \bibinfo {author} {\bibfnamefont {A.~P.}\
  \bibnamefont {Seitsonen}}, \bibinfo {author} {\bibfnamefont {A.}~\bibnamefont
  {Smogunov}}, \bibinfo {author} {\bibfnamefont {I.}~\bibnamefont {Timrov}},
  \bibinfo {author} {\bibfnamefont {T.}~\bibnamefont {Thonhauser}}, \bibinfo
  {author} {\bibfnamefont {P.}~\bibnamefont {Umari}}, \bibinfo {author}
  {\bibfnamefont {N.}~\bibnamefont {Vast}}, \bibinfo {author} {\bibfnamefont
  {X.}~\bibnamefont {Wu}},\ and\ \bibinfo {author} {\bibfnamefont
  {S.}~\bibnamefont {Baroni}},\ }\bibfield  {title} {\bibinfo {title} {Advanced
  capabilities for materials modelling with {Quantum} {ESPRESSO}},\ }\href
  {https://doi.org/10.1088/1361-648X/aa8f79} {\bibfield  {journal} {\bibinfo
  {journal} {Journal of Physics: Condensed Matter}\ }\textbf {\bibinfo {volume}
  {29}},\ \bibinfo {pages} {465901} (\bibinfo {year} {2017})}\BibitemShut
  {NoStop}%
\bibitem [{\citenamefont {Mostofi}\ \emph {et~al.}(2014)\citenamefont
  {Mostofi}, \citenamefont {Yates}, \citenamefont {Pizzi}, \citenamefont {Lee},
  \citenamefont {Souza}, \citenamefont {Vanderbilt},\ and\ \citenamefont
  {Marzari}}]{Mostofi2014}%
  \BibitemOpen
  \bibfield  {author} {\bibinfo {author} {\bibfnamefont {A.~A.}\ \bibnamefont
  {Mostofi}}, \bibinfo {author} {\bibfnamefont {J.~R.}\ \bibnamefont {Yates}},
  \bibinfo {author} {\bibfnamefont {G.}~\bibnamefont {Pizzi}}, \bibinfo
  {author} {\bibfnamefont {Y.-S.}\ \bibnamefont {Lee}}, \bibinfo {author}
  {\bibfnamefont {I.}~\bibnamefont {Souza}}, \bibinfo {author} {\bibfnamefont
  {D.}~\bibnamefont {Vanderbilt}},\ and\ \bibinfo {author} {\bibfnamefont
  {N.}~\bibnamefont {Marzari}},\ }\bibfield  {title} {\bibinfo {title} {An
  updated version of wannier90: A tool for obtaining maximally-localised
  wannier functions},\ }\href {https://doi.org/10.1016/j.cpc.2014.05.003}
  {\bibfield  {journal} {\bibinfo  {journal} {Computer Physics Communications}\
  }\textbf {\bibinfo {volume} {185}},\ \bibinfo {pages} {2309} (\bibinfo {year}
  {2014})}\BibitemShut {NoStop}%
\bibitem [{\citenamefont {Mostofi}\ \emph {et~al.}(2008)\citenamefont
  {Mostofi}, \citenamefont {Yates}, \citenamefont {Lee}, \citenamefont {Souza},
  \citenamefont {Vanderbilt},\ and\ \citenamefont {Marzari}}]{Mostofi2008}%
  \BibitemOpen
  \bibfield  {author} {\bibinfo {author} {\bibfnamefont {A.~A.}\ \bibnamefont
  {Mostofi}}, \bibinfo {author} {\bibfnamefont {J.~R.}\ \bibnamefont {Yates}},
  \bibinfo {author} {\bibfnamefont {Y.-S.}\ \bibnamefont {Lee}}, \bibinfo
  {author} {\bibfnamefont {I.}~\bibnamefont {Souza}}, \bibinfo {author}
  {\bibfnamefont {D.}~\bibnamefont {Vanderbilt}},\ and\ \bibinfo {author}
  {\bibfnamefont {N.}~\bibnamefont {Marzari}},\ }\bibfield  {title} {\bibinfo
  {title} {wannier90: {A} tool for obtaining maximally-localised {Wannier}
  functions},\ }\href {https://doi.org/10.1016/j.cpc.2007.11.016} {\bibfield
  {journal} {\bibinfo  {journal} {Computer Physics Communications}\ }\textbf
  {\bibinfo {volume} {178}},\ \bibinfo {pages} {685} (\bibinfo {year}
  {2008})}\BibitemShut {NoStop}%
\bibitem [{\citenamefont {Allen}(1972)}]{allen_neutron_1972}%
  \BibitemOpen
  \bibfield  {author} {\bibinfo {author} {\bibfnamefont {P.~B.}\ \bibnamefont
  {Allen}},\ }\bibfield  {title} {\bibinfo {title} {Neutron {Spectroscopy} of
  {Superconductors}},\ }\href {https://doi.org/10.1103/PhysRevB.6.2577}
  {\bibfield  {journal} {\bibinfo  {journal} {Physical Review B}\ }\textbf
  {\bibinfo {volume} {6}},\ \bibinfo {pages} {2577} (\bibinfo {year}
  {1972})}\BibitemShut {NoStop}%
\bibitem [{\citenamefont {Allen}\ and\ \citenamefont
  {Dynes}(1975)}]{Allen1975}%
  \BibitemOpen
  \bibfield  {author} {\bibinfo {author} {\bibfnamefont {P.~B.}\ \bibnamefont
  {Allen}}\ and\ \bibinfo {author} {\bibfnamefont {R.~C.}\ \bibnamefont
  {Dynes}},\ }\bibfield  {title} {\bibinfo {title} {Transition temperature of
  strong-coupled superconductors reanalyzed},\ }\href
  {https://doi.org/10.1103/PhysRevB.12.905} {\bibfield  {journal} {\bibinfo
  {journal} {Physical Review B}\ }\textbf {\bibinfo {volume} {12}},\ \bibinfo
  {pages} {905} (\bibinfo {year} {1975})}\BibitemShut {NoStop}%
\bibitem [{\citenamefont {Wu}\ \emph {et~al.}(2018{\natexlab{b}})\citenamefont
  {Wu}, \citenamefont {Zhang}, \citenamefont {Song}, \citenamefont {Troyer},\
  and\ \citenamefont {Soluyanov}}]{Wu2018a}%
  \BibitemOpen
  \bibfield  {author} {\bibinfo {author} {\bibfnamefont {Q.}~\bibnamefont
  {Wu}}, \bibinfo {author} {\bibfnamefont {S.}~\bibnamefont {Zhang}}, \bibinfo
  {author} {\bibfnamefont {H.-F.}\ \bibnamefont {Song}}, \bibinfo {author}
  {\bibfnamefont {M.}~\bibnamefont {Troyer}},\ and\ \bibinfo {author}
  {\bibfnamefont {A.~A.}\ \bibnamefont {Soluyanov}},\ }\bibfield  {title}
  {\bibinfo {title} {Wanniertools: An open-source software package for novel
  topological materials},\ }\href {https://doi.org/10.1016/j.cpc.2017.09.033}
  {\bibfield  {journal} {\bibinfo  {journal} {Computer Physics Communications}\
  }\textbf {\bibinfo {volume} {224}},\ \bibinfo {pages} {405} (\bibinfo {year}
  {2018}{\natexlab{b}})}\BibitemShut {NoStop}%
\bibitem [{\citenamefont {Yu}\ \emph {et~al.}(2011)\citenamefont {Yu},
  \citenamefont {Qi}, \citenamefont {Bernevig}, \citenamefont {Fang},\ and\
  \citenamefont {Dai}}]{Yu2011}%
  \BibitemOpen
  \bibfield  {author} {\bibinfo {author} {\bibfnamefont {R.}~\bibnamefont
  {Yu}}, \bibinfo {author} {\bibfnamefont {X.~L.}\ \bibnamefont {Qi}}, \bibinfo
  {author} {\bibfnamefont {A.}~\bibnamefont {Bernevig}}, \bibinfo {author}
  {\bibfnamefont {Z.}~\bibnamefont {Fang}},\ and\ \bibinfo {author}
  {\bibfnamefont {X.}~\bibnamefont {Dai}},\ }\bibfield  {title} {\bibinfo
  {title} {Equivalent expression of {Z2} topological invariant for band
  insulators using the non-{Abelian} {Berry} connection},\ }\href
  {https://doi.org/10.1103/PhysRevB.84.075119} {\bibfield  {journal} {\bibinfo
  {journal} {Physical Review B}\ }\textbf {\bibinfo {volume} {84}},\ \bibinfo
  {pages} {075119} (\bibinfo {year} {2011})}\BibitemShut {NoStop}%
\bibitem [{\citenamefont {Kane}\ and\ \citenamefont {Mele}(2005)}]{Kane2005}%
  \BibitemOpen
  \bibfield  {author} {\bibinfo {author} {\bibfnamefont {C.~L.}\ \bibnamefont
  {Kane}}\ and\ \bibinfo {author} {\bibfnamefont {E.~J.}\ \bibnamefont
  {Mele}},\ }\bibfield  {title} {\bibinfo {title} {{Z2} {Topological} {Order}
  and the {Quantum} {Spin} {Hall} {Effect}},\ }\href
  {https://doi.org/10.1103/PhysRevLett.95.146802} {\bibfield  {journal}
  {\bibinfo  {journal} {Physical Review Letters}\ }\textbf {\bibinfo {volume}
  {95}},\ \bibinfo {pages} {146802} (\bibinfo {year} {2005})}\BibitemShut
  {NoStop}%
\bibitem [{\citenamefont {Chen}\ \emph {et~al.}(2023)\citenamefont {Chen},
  \citenamefont {Shen}, \citenamefont {Feng}, \citenamefont {Liu},
  \citenamefont {Liu},\ and\ \citenamefont {Wu}}]{Chen_2023}%
  \BibitemOpen
  \bibfield  {author} {\bibinfo {author} {\bibfnamefont {W.}~\bibnamefont
  {Chen}}, \bibinfo {author} {\bibfnamefont {T.}~\bibnamefont {Shen}}, \bibinfo
  {author} {\bibfnamefont {Y.}~\bibnamefont {Feng}}, \bibinfo {author}
  {\bibfnamefont {C.}~\bibnamefont {Liu}}, \bibinfo {author} {\bibfnamefont
  {X.}~\bibnamefont {Liu}},\ and\ \bibinfo {author} {\bibfnamefont
  {Y.}~\bibnamefont {Wu}},\ }\bibfield  {title} {\bibinfo {title} {Sensitive
  biosensing based on a d-type photonic crystal fiber},\ }\href
  {https://doi.org/10.1088/1402-4896/acf961} {\bibfield  {journal} {\bibinfo
  {journal} {Physica Scripta}\ }\textbf {\bibinfo {volume} {98}},\ \bibinfo
  {pages} {105531} (\bibinfo {year} {2023})}\BibitemShut {NoStop}%
\bibitem [{\citenamefont {Akiba}\ and\ \citenamefont
  {Kobayashi}(2023)}]{akiba2023phonon}%
  \BibitemOpen
  \bibfield  {author} {\bibinfo {author} {\bibfnamefont {K.}~\bibnamefont
  {Akiba}}\ and\ \bibinfo {author} {\bibfnamefont {T.~C.}\ \bibnamefont
  {Kobayashi}},\ }\bibfield  {title} {\bibinfo {title} {Phonon-mediated
  superconductivity in the sb square-net compound lacusb 2},\ }\href@noop {}
  {\bibfield  {journal} {\bibinfo  {journal} {Physical Review B}\ }\textbf
  {\bibinfo {volume} {107}},\ \bibinfo {pages} {245117} (\bibinfo {year}
  {2023})}\BibitemShut {NoStop}%
\bibitem [{\citenamefont {Poncé}\ \emph {et~al.}(2016)\citenamefont {Poncé},
  \citenamefont {Margine}, \citenamefont {Verdi},\ and\ \citenamefont
  {Giustino}}]{Ponce2016}%
  \BibitemOpen
  \bibfield  {author} {\bibinfo {author} {\bibfnamefont {S.}~\bibnamefont
  {Poncé}}, \bibinfo {author} {\bibfnamefont {E.~R.}\ \bibnamefont {Margine}},
  \bibinfo {author} {\bibfnamefont {C.}~\bibnamefont {Verdi}},\ and\ \bibinfo
  {author} {\bibfnamefont {F.}~\bibnamefont {Giustino}},\ }\bibfield  {title}
  {\bibinfo {title} {{EPW}: {Electron}–phonon coupling, transport and
  superconducting properties using maximally localized {Wannier} functions},\
  }\href {https://doi.org/10.1016/j.cpc.2016.07.028} {\bibfield  {journal}
  {\bibinfo  {journal} {Computer Physics Communications}\ }\textbf {\bibinfo
  {volume} {209}},\ \bibinfo {pages} {116} (\bibinfo {year}
  {2016})}\BibitemShut {NoStop}%
\bibitem [{\citenamefont {Soluyanov}\ and\ \citenamefont
  {Vanderbilt}(2011)}]{Soluyanov2011a}%
  \BibitemOpen
  \bibfield  {author} {\bibinfo {author} {\bibfnamefont {A.~A.}\ \bibnamefont
  {Soluyanov}}\ and\ \bibinfo {author} {\bibfnamefont {D.}~\bibnamefont
  {Vanderbilt}},\ }\bibfield  {title} {\bibinfo {title} {Wannier representation
  of {Z}\_2 topological insulators},\ }\href
  {https://doi.org/10.1103/PhysRevB.83.035108} {\bibfield  {journal} {\bibinfo
  {journal} {Physical Review B}\ }\textbf {\bibinfo {volume} {83}},\ \bibinfo
  {pages} {035108} (\bibinfo {year} {2011})},\ \bibinfo {note} {arXiv:1009.1415
  [cond-mat]}\BibitemShut {NoStop}%
\end{thebibliography}%
\end{document}